\documentclass[galaxies,article,accept,oneauthor,10pt,a4paper]{mdpi}

\firstpage{1} 
\articlenumber{x}
\doinum{10.3390/------}
\pubvolume{xx}
\pubyear{2016}
\copyrightyear{2016}
\externaleditor{Academic Editor: name}
\history{Received: date; Accepted: date; Published: date}

\Title{On the dwarf galaxies rotation curves diversity problem}

\AuthorNames{Antonino Del Popolo, Morgan Le Delliou, Xiguo Lee}

\Author{Antonino Del Popolo $^{1,2,3,\dagger}$, Morgan Le Delliou $^{4,5,\ddagger}$, Xiguo Lee $^{4,\mathsection}$}
\AuthorNames{Antonino Del Popolo, Morgan Le Delliou, Xiguo Lee}

\address{$^{1}$ Dipartimento di Fisica e Astronomia, University Of Catania
, Viale Andrea Doria 6, 95125
Catania, Italy
 \quad  \\
$^{2}$ Institute of Modern Physics, Chinese Academy of Sciences, Post Oce Box31, Lanzhou 730000, Peoples Republic of China \quad  \\
$^{3}$ INFN sezione di Catania,  Via S. Sofia 64, I-95123 Catania, Italy \quad  \\
$^{4}$ Institute of Theoretical Physics, Physics Department, Lanzhou University, No.222, South Tianshui Road, Lanzhou, Gansu 730000, P R China 
 \\
$^{5}$ Centro de Astronomia e Astrofísica da Universidade de Lisboa, Faculdade de Ciências, Ed. C8, Campo Grande, 1769-016 Lisboa, Portugal\\
}

\corres{Correspondence: delliou@lzu.edu.cn}

\firstnote{adelpopolo@oact.inaf.it} 
\secondnote{delliou@lzu.edu.cn}
\thirdnote{xgl@impcas.ac.cn}

\abstract{In this paper, we show how baryonic physics can solve the problem of the striking diversity in dwarf galaxies rotation curves shapes. To this aim, we compare the distribution of galaxies of the SPARC sample, in the plane $V_{\rm 2 kpc}$-$V_{\rm Rlast }$ (being 
$V_{\rm 2kpc}$ the galaxy rotation velocity at $2$ kpc, and $V_{\rm Rlast}$ that outermost one) with that of galaxies that we simulated taking account of baryonic effects.
The scatter in the rotation curves in the $V_{\rm 2 kpc}$-$V_{\rm Rlast }$ plane, and  the trend of the SPARC sample's, and our simulated galaxies', distribution is in good agreement. 
The solution of the "diversity" problem lies in the ability of baryonic process to produce non self-similar haloes, contrary to DM-only simulations.  
We show also that baryonic effects can reproduce the rotation curves of galaxies like IC2574 characterized by a slow rising with radius. 
A solution to the diversity problem can be obtained taking appropriately into account the baryon physics effects.}


\begin{document}



\section{Introduction}

Among the predictions of the $\Lambda$CDM paradigm, agreeing very well with a plethora of observations \citep{DelPopolo2007,Komatsu2011,DelPopolo2013,Story2013,Das2014,DelPopolo2014,Planck2016_XIII}\footnote{
Here, we should recall that in addition to some small scale problems, the $\Lambda$CDM paradigm entails two unsolved problems: the cosmic coincidence problem, and the cosmological constant problem 
\citep{Weinberg1989,Astashenok2012}.}, appears the forecast that dark matter (DM) haloes have a cuspy density profile, with $\rho \propto r^{-1}$ {\citep{Navarro1997}}, close to the halo center, given by the so called Navarro-Frenk-White (NFW) profile 
\cite{Navarro1997}. More recent findings confirm this result, however with a lower slope for the cusp {\citep{Gao2008,Navarro2010}}. The rotation curves (RCs) of dwarf, and low brightness galaxies (LSB) are usually characterized by a more gentle increase than predicted by the NFW profile . This problem was noticed for the first time more than two decades ago \citep{Moore1994,Flores1994}, studied in many other papers {\citep{Burkert1995,deBlok2003,Governato2010,KuziodeNaray2011,Oh2011b,Cardone2011a,Cardone2011b,Cardone2012,DelPopolo2012a,DelPopolo2012b,
DelPopolo2012c,DelPopolo2013d,DiCintio2014,DelPopolo2014a,DelPopolo2014b,DelPopolo2014d,Polisensky2015}, and recently it was shown to be present even at cluster of galaxies scale \citep{Sand2002,Sand2004,Newman2013a,Newman2013b}}.

From a more general point of view, the cusp/core problem is better defined in terms of the excess of DM predicted in the inner parts of the galaxies compared with the observed inner slope, and it can be connected to the Too-Big-To-Fail problem {\citep{BoylanKolchin2011,Oman2015,Papastergis2015,DelPopolo2017a}}.  

While dwarf galaxies usually have cored profiles, a more detailed study shows a significant spread in their RCs, and the existence of cuspy dwarfs
{\citep{Simon2005,Oh2011b,Adams2014}}.

Despite disagreement on the above discrepancy, the inner profiles of several dwarf galaxies or LSBs are clearly not always flat \citep[e.g.][]{Simon2005,Oman2015}, and \citep{deBlok2008} noticed in 
the THINGS sample a clear mass dependence of inner profiles. \citep{Ricotti2003} first predicted such mass dependence {by using DM-only simulations}.   Inner slope dependence on the halo 
mass and on the ratio $(M_{gas}+M_*)/M_{total}$, where $M_*$ is the stellar mass, was shown in more recent studies {\citep{DelPopolo2010,DelPopolo2011,DiCintio2014}}.

In addition, currently, different techniques (e.g., spherical Jeans equation, multiple stellar populations techniques, Schwarzschild modeling) applied to those or similar objects sometimes give different 
results (e.g., the cuspy profile found by \cite{Breddels2013} in Fornax, while 
{\citep{Walker2011,Battaglia2008,Agnello2012,Genina2018}}, 
found a core). Finally, discrepancies are evident for larger objects than those MW satellites (e.g., \cite{Simon2003} found $-0.17< \alpha <-0.01$ in the case of NGC2976, while
\citep{Adams2012}
found $\alpha=-0.90 \pm 0.15$, and
\citep{Adams2014} found $\alpha=-0.53 \pm 0.14$
if stars were used to trace the potential, or 
$\alpha=-0.30 \pm 0.18$
if it was the gas).

From the studies discussed above, and several others, results show the existence of a range of profiles, and that no agreement on the exact dark matter slopes {\citep{Simon2005,Oh2011b,Adams2014}}
distribution {can be reached} based on morphologies, despite current improvements in kinematic maps.
{Namely, despite using the most recent and accurate data (kinematic maps), it is very complicated to distinguish between cusp and cores even in galaxies of same morphology, and in some case even for the same object. This problem is more evident in dwarf galaxies (see the above citations).}

Recently, \cite{Oman2015} quantified this diversity in dwarf galaxies RCs. They compared the circular velocity at 2 kpc $V_{2 kpc}$, given a fixed maximum in the circular velocity, 
$V_{max}$. The scatter in $V_{2 kpc}$, in the $V_{max}$ range 50-250 $km/s$, spans a factor 3-4. 

Several authors proposed solutions to that problem, almost all relying on core formation process due to supernovae feedback (hereafter SNFM). At this point we want to emphasize that despite the majority of studies 
converging on the idea that baryon physics leads to the formation of cores \citep{DelPopolo2009,Governato2010,DelPopolo2010,DelPopolo2012a,DelPopolo2012b,Onorbe2015,Read2016a}, some studies arrive at the opposite conclusion 
\citep{Vogelsberger2014,Gonzalez-Samaniego2014,Sawala2015,Schaller2015}. While this 
disagreement can be due to different physical processes included in the simulations, it motivates one to be more careful in accepting simulation results. As for the diversity problem, it brings about one question: why wasn't the 
diversity problem seen and solved by the large number of hydro-simulations run in the last decade, and especially in the past years, before \cite{Oman2015} pointed it out, if the galaxies they formed were as realistic as claimed? 
Actually, some years before \cite{Oman2015}, the problem was discussed and solved by means of baryon physics, using a semi-analytical model in \cite{DelPopolo2012a}. That core formation model, differently from the SNFM, is related to the exchange of energy and angular momentum (AM) between gas clumps and DM through dynamical friction (dubbed dynamical friction from baryonic clumps model (DFBC))\citep{ElZant2001,ElZant2004,Ma2004,RomanoDiaz2008,RomanoDiaz2009,DelPopolo2009,DelPopolo2010,Cole2011,
Inoue2011,DelPopolo2012a,DelPopolo2012b,Nipoti2015,DelPopolo2016a,DelPopolo2016b}.

However, after the \cite{Oman2015} analysis, several authors claimed the problem could be solved by the same simulations, based on the SNFM, that were previously blind to it. In particular, \cite{Read2016b} showed that similar simulations can solve the diversity problem with cores formation by baryons, while \cite{Oman2015} could not. Even in the case that the core formation scenario through supernovae feedback can solve the diversity problem, such scenario encounters serious difficulties to explain the structure of objects like IC2574 {\citep{Oman2015,Oman2016,Creasey2017}}, which displays a core extending to 8 kpc, where there is no star (the ICG2574 half-mass radius is 5 kpc). Problems like this prompted 
\cite{Creasey2017} to explore a solution based on self-interacting dark matter (SIDM): they found that the SIDM alone cannot solve the problem and thus somehow some baryonic physics must be introduced. 

In this paper, the distribution of galaxies produced by the DFBC in the $V_{\rm 2 kpc}$-$V_{\rm Rlast }$, where $V_{\rm Rlast}$ is the outermost radius,  
will be compared, following \cite{Oman2015}, to the SPARC data \citep{Lelli2016a}, a collection of high quality RCs of nearby 
galaxies. 

We expect the mass dependence of the inner structure of the galaxies, as shown in \cite{DelPopolo2010,DelPopolo2016a}, to give rise to a scatter in the $V_{2 kpc}$-$V_{\rm Rlast}$ plane. Such scatter is not possible in the CDM scenario, producing self-similar DM haloes. 

In the first Sec.~ 2
the model and observations will be briefly detailed, and their confrontation discussed in Sec.~ 3
before concluding remarks in 
Sec.~ 4


\section{Model and comparison with observations}\label{sec:Modelobs}
\textnormal{
Our study of the diversity problem involved a subsample of the SPARC sample \citep{Lelli2016a}, which is a collection of nearby galaxies high-quality rotation curves, to determine $V_{2 kpc}$, and $V_{\rm Rlast}$. The subsample characteristics are described in Sec. \ref{sec:ObsDat}}\textnormal{.}

\textnormal{We then simulated 100 galaxies with the 
DFBC model \citep{DelPopolo2009,DelPopolo2016b}, with similar characteristics to our SPARC subsample, and with $M_*=10^7-10^{11} M_{\odot}$, in a $\Lambda$CDM cosmology according to the \cite{Planck2014} 
parameters. }

Finally, we compared the SPARC subsample and simulated $V_{2 kpc}$, $V_{\rm Rlast}$, as summarized in the following.

\subsection{Model}

The model simulating galaxy formation we used has been described in several papers \citep{DelPopolo2009,DelPopolo2012a,DelPopolo2012b,DelPopolo2016a,DelPopolo2016b}.
It provides a highly improved spherical collapse models from that described by  
\citep{Gunn1972,Bertschinger1985,Hoffman1985,Ryden1987,Ascasibar2004,Williams2004}, and includes the effects of dark energy \citep{DelPopolo2013a,DelPopolo2013b,DelPopolo2013c},  random angular momentum \citep[e.g.,][]{Ryden1987,Williams2004} produced by the random motions arising in the collapse phase, ordered angular momentum 
\citep[e.g.,][]{Ryden1988,DelPopolo1997,DelPopolo2000} arising from tidal torques, adiabatic contraction \citep[e.g.,][]{Blumenthal1986,Gnedin2004, Klypin2002,Gustafsson2006}, gas and stellar clumps interactions with DM through dynamical friction 
\citep{ElZant2001,ElZant2004,Ma2004,RomanoDiaz2008,RomanoDiaz2009,DelPopolo2009,Cole2011,Inoue2011, Nipoti2015}, gas cooling, star formation, photoionization, supernova, and AGN feedback 
\citep[see the following]{DeLucia2008,Li2010,Martizzi2012}. 

It follows perturbations of diffuse gas (baryons) and DM, which will give rise to the proto-structure, from the linear to non-linear phase, through turn-around and collapse. Baryons fraction is set to the 
"universal baryon fraction" $f_b = 0.17 \pm 0.01$ \citep{Komatsu2009}
\citep[0.167 in][]{Komatsu2011}. The baryons collapse compresses 
DM (adiabatic compression), steepening the DM profile \citep{Blumenthal1986,Gnedin2004,Gustafsson2006}.

If a DM particle is located at a given radius $r< r_i$ 
\begin{equation}
r_i M_i (r_i)=r \left [ M_b( r) +M_{dm} ( r) \right] 
\label{eq:ad1}
\end{equation}
\citep{Blumenthal1986,Ryden1988,Flores1993}
where $M_i (r_i)$ is the initial dark halo distribution, then $M_b$ the final mass distribution of baryons 
(i.e. for example, an exponential disk for spirals, or a Hernquist configuration
(\citep{Keeton2001,Treu2002} for elliptical galaxies), $r$ the final radius, and $M_{dm}$ the final DM 
distribution, are obtained through solving iteratively Eq. \ref{eq:ad1} \citep{Spedicato2003}.
This model can be improved to better reproduce numerical simulations by assuming conservation of the product of the radius by the inside mass for that orbit-averaged radius \citep{Gnedin2004}.

Radiative processes form baryons clumps, in turn collapsing towards the center of the galaxy and forming stars. Clumps formation, their life-time, and observation are discussed in \cite{DelPopolo2016b}.

Dynamical friction (DF) between baryons and DM transfers energy and angular momentum (AM) to the DM component \citep{ElZant2001, ElZant2004}. This gives rise to a predominantly outwards motion of DM particles, reducing the central 
density, and transforming the cusp into a core. \citep[][Appendix D]{DelPopolo2009} describes how DF is taken into account, inserting the DF force in the equation of motion \citep[Eq. A14,][]{DelPopolo2009} to affects structure formation.

That mechanism, flattening the cusp, is amplified in the case of rotationally supported (spiral) galaxies through the ordered AM, $L$, acquired by the proto-structure through tidal interactions with neighbors 
\citep{Peebles1969,White1984,Ryden1988,Eisenstein1995}, 
and by random AM, $j$, originating by the random motions arising in the collapse phase \citep{Ryden1987}. 

The ``ordered" AM is calculated from evaluating the torque $\tau(r)$, and integrating it over time 
(\cite{Ryden1988}, equation 35; see also Sect. C2 of \cite{DelPopolo2009}).
``Random" AM is taken into account by assigning a specific angular momentum at turnaround (for details see Appendix C2 of \citep{DelPopolo2009}).

A classical cooling flow served as gas cooling mechanism \citep[e.g.,][]{White1991} \citep[see Sect.~2.2.2 of][]{Li2010}. The inclusion by \cite{DeLucia2008} and \cite[Sect.~2.2.2 and~2.2.3]{Li2010} of star formation, 
reionization and supernovae feedback were replicated.
Following \cite{Li2010}, reionization reduces the baryon fraction by
\begin{equation}
 f_{\rm b, halo}(z,M_{\rm vir})=\frac{f_{\rm b}}{[1+0.26 M_{\rm F}(z)/M_{\rm vir}]^3}\;,
\end{equation}
with the virial mass  $M_{\rm vir}$ and $M_{\rm F}$ is the "filtering mass" \citep*[see][]{Kravtsov2004}. We take the reionization redshift in the range 11.5-15. Our treatment of supernovae feedback also follows \cite{Croton2006}. 
In that stage, each supernova explosion expels gas in successive events, lowering stellar density. The smallest clumps are  destroyed by feedback soon after stars are formed from a small part of their mass \citep{Nipoti2015}. 

AGN quenching becomes important for masses $\simeq 6 \times 10^{11} M_{\odot}$ \citep{Cattaneo2006}. Its feedback was taken into account modifying \cite{Booth2009} model as in Sec. 2.3 of 
\cite{Martizzi2012}, by forming a 
Super-Massive-Black-Hole (SMB) when the star density exceeds $2.4 \times 10^6 M_{\odot}/kpc^3$, then accreting mass into it.

Our model demonstrated its robustness in several ways:   
\renewcommand{\theenumi}{\alph{enumi}}\begin{enumerate}
 \item cusp flattening from DM heating by collapsing baryonic clumps is in agreement with previous studies \citep{ElZant2001,ElZant2004,RomanoDiaz2008,RomanoDiaz2009,Cole2011,Inoue2011,Nipoti2015}. 
{\citep{DelPopolo2011}, Fig 4, shows a comparison of our model with the SPH simulations of \cite{Governato2010}}.
\item galaxies density profiles correct shape \citep{DelPopolo2009,DelPopolo2009a}, this before the
\cite{Governato2010,Governato2012}  
SPH simulations, and correct clusters density profiles \citep{DelPopolo2012b} were 
 predicted, and a series of correlations in clusters' observations \citep{Newman2013a,Newman2013b} were reobtained \cite{DelPopolo2014c}. {Notice that concerning  
correlations in clusters of galaxies, in \cite{DelPopolo2014}, Fig. 2-5, we showed a comparison with 
\citep{Newman2013b} observations. }
\item inner slope dependence on halo mass \cite{DelPopolo2010}, and on the total baryonic content to total mass ratio \cite{DelPopolo2012b} were predicted, in agreement with \cite{DiCintio2014}. In
 addition to these dependence, the inner slope was also found to depend on the angular momentum
\cite{DelPopolo2012b}.  {\citep{DelPopolo2016a,DelPopolo2016b} is showing in Fig. 1 a comparison of the change of the inner slope with mass with \citep{DiCintio2014} simulations. Moreover Figs. 4, 5 in \citep{DelPopolo2016a,DelPopolo2016b}, shows a comparison of: the Tully-Fisher, Faber-Jackson, 
 $M_{Star}-M_{halo}$ relationship, with simulations.}
Finally, the correct DM profile 
inner slope dependence on the halo mass is explained over 6 order of magnitudes in halo mass, from dwarves to clusters \citep{DelPopolo2009,DelPopolo2010,DelPopolo2012a,DelPopolo2012b,DelPopolo2014c}, a range that no other model can achieve.
\end{enumerate}

\begin{figure}[!t]
 \centering
 \includegraphics[scale=0.9]{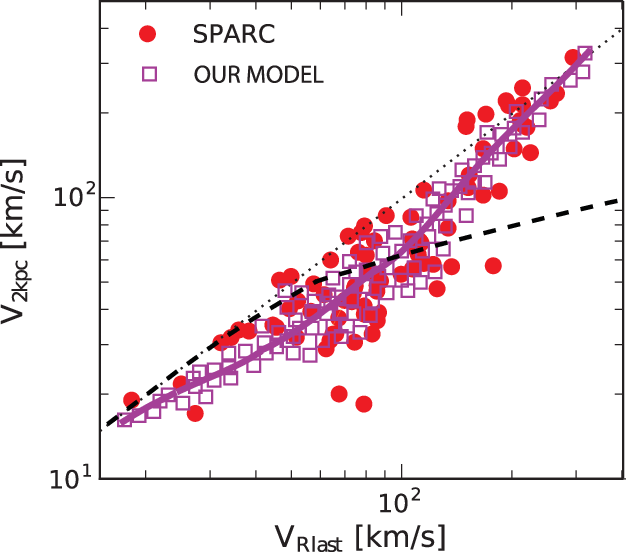}
 \includegraphics[scale=0.9]{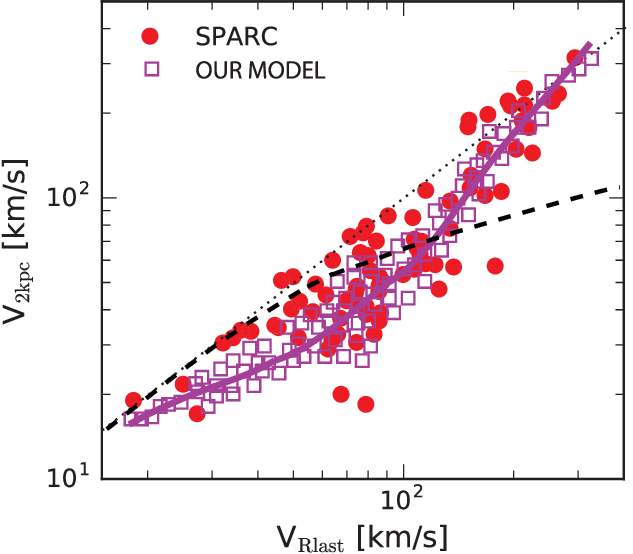}
 \par
 \caption[]{Comparison of the prediction from our model with 
 a selection of SPARC data, using (a) (top panel) the spherical circular velocity approximation; (b) (bottom panel) the disk plane 
 cylindrical gravitational potential approximation.
 The full dots represent the SPARC sample, the open violet squares our simulated galaxies, the violet line represents the mean trend line, while the dashed line the expectation if a NFW profile described all haloes.
 }
 \label{fig:NIHAOus}
\end{figure}

\subsection{Observational data}\label{sec:ObsDat}

The choice we made of galaxies observations was a subsample extracted from the Spitzer Photometry \& Accurate Rotation Curves (SPARC) \citep{Lelli2016a}. Spanning large ranges in morphologies, surface brightnesses, and luminosities,
and presenting new surface photometry at 3.6 $\mu$m and high-quality rotation curves from previous $\rm
HI/H\alpha$ studies, the entire sample contains 175 nearby disk galaxies. To minimizes the scatter in the baryonic Tully-Fisher 
relation, and following \cite{Lelli2016b}, we assumed a mass-to-light ratio $\Upsilon_*=0.5 M_{\odot}/L_{\odot}$. 

{
For galaxies, not having a measured point at 2 kpc
we used an interpolation between close bins.
}

We also applied to the sample the conditions  
\begin{enumerate}
 \item selecting stellar masses similar to our simulated galaxies, and 
 \item selecting galaxies with inclination $>45^{\rm o}$ which are the most reliable RC data.
\end{enumerate}

\section{Results}\label{sec:Results}

Once the observed sample and model simulated galaxies populations were determined, we compared their 
location in the plane $V_{\rm 2kpc}-V_{\rm Rlast}$. {We defined $R_{\rm last}$ from the relation 
$\log(R_{\rm last}/kpc)=0.31 \times \log(M_\star/M_\odot)-1.8$, that only slightly differs from the Santos-Santos case \cite[, Fig. 1]{Santos-Santos2017}.  }

In Fig.~\ref{fig:NIHAOus}, we compared the SPARC data 
with the result of our model.
The models interpretation of circular velocity are based on the 3D spherical circular velocity approximation ($V_{\rm circ-spherical}=\sqrt{\frac{GM(<r)}{r}}$, with $r$ the 3D radius and $M(<r)$ its enclosed mass) in 
Fig.~\ref{fig:NIHAOus}a (top panel of Fig. 1) and calculated using the disk plane cylindrical gravitational potential ($V_{\rm circ-potential}=\sqrt{R\frac{\partial \Phi}{\partial R}_{z=0}}$, with $R$ the cylindrical radius and $\Phi$ the disk potential, restricted in the galactic plane $z=0$) in Fig.~\ref{fig:NIHAOus}b (bottom panel of Fig. 1). 
In both panels of Fig.~\ref{fig:NIHAOus},  
the full dots represent the SPARC sample. The open violet squares mark our simulated galaxies,
the violet line represents the mean trend line, while the dashed line the expectation if a NFW profile described all haloes.
The plot shows the region occupied by the galaxies
distributions predicted by our model are in much better agreement with the SPARC galaxy distribution and its scatter than the output of \cite{Oman2015}. 
%
%

{Notice, that the robustness of the quoted results are independent of the choice of inner velocity (here 2 kpc) as we verified, considering smaller radii(1 kpc), and as was checked by \cite[Fig.3]{Santos-Santos2017} }

Denoting as `outliers' SPARC galaxies having $V_{\rm 2kpc}$ outside the $\pm 3\sigma$ range determined with respect to our model trend line, there are two outliers,
namely: IC2574, UCG05750.
The circular-potential velocity definition employed in Fig.~\ref{fig:NIHAOus}b lowers the trend line, especially in the maximum feedback region, and gives rise to a small increase in scatter. This reintegrates, on one hand, the outliers IC2574, and UGC05750.
{Moreover, the error-bars on SPARC galaxies were not shown in the figures, in order to obtain a cleaner plot. Taking account of errors, on average of $5-10$ km/s for SPARC, renders even more evident that the galaxies are not outliers. }

\begin{figure}[tbp]
 \centering
 \includegraphics[scale=0.45]{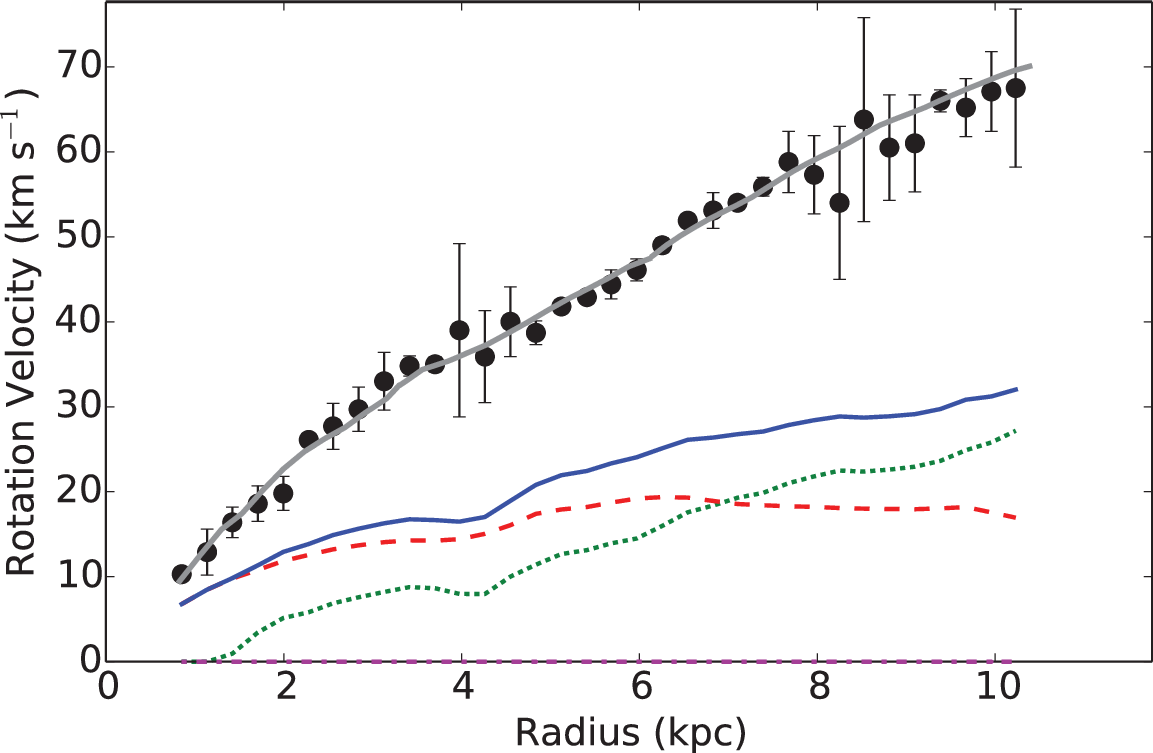}
 \par
 \caption[]{Rotation curve of IC2574 compared with our results. The SPARC rotation curve of IC2574 is presented as dots with error bars. Our simulations result yields the grey solid line, including the RCs from its stars 
 (dotted green line), and gas disk (dashed red line) components, themselves summed into the total baryonic mass (blue solid line).}
 \label{fig:RCvsModelIC2574}
\end{figure}

The successful reproduction of a similar distribution and scatter to SPARC galaxies by our model 
reflects the ability to model the total and stellar mass dependence of galaxies DM density profile and 
RCs, as shown in \cite{DelPopolo2016a}: {galaxies with $M_* \simeq 10^8 M_{\odot}$ are cored with a very flat profile and inner slope $\alpha \simeq 0$, while galaxies with $M_*<10^8$ tend to be more cuspy (larger $\alpha$) as the feedback mechanism efficiency, producing the cores, is decreased for lack of baryon clumps.} Conversely, galaxies with $M_*>10^8$ also tend to be cuspier. This time the cuspy profiles are produced by
the presence of a larger number of stars, which deepens the galactic potential well, and opposes the SNFM mechanism.

It is important to notice that our galaxies follow the SPARC trend at 
$V_{\rm Rlast} \geq 150 km/s$,
thanks to our accounting of AGN feedback that counteracts baryons cooling, and modifies star formation. In addition, tidal interaction in our model makes it more environment dependent than 
several hydro-dynamic simulations, in which galaxies are usually isolated.

A recent paper of \citep{Santos-Santos2017} shows similar results. We want to stress, that the method used in the two papers is totally different: simulations (\citep{Santos-Santos2017}), vs a semi-analytic method (this paper). The explanation of the "diversity problem" is related to the interaction of baryon clumps with dark matter, and not due to Supernovae Feedback. Our results are dependent from environment (galaxies are not isolated), and the effect of AGN feedback, differently from \citep{Santos-Santos2017}. Those effects change the distribution of galaxies in the $V_{\rm 2kpc}$-$V_{\rm Rlast}$ plane. There are also differences in the SPARC sample used. In that of \citep{Santos-Santos2017} some SPARC galaxies, outliers in their case, were not plotted by the authors. 

As previously discussed, IC2574, and UGC05750 are no more outliers, once calculating $V_{\rm 2kpc}$ in the galactic plane, as seen in the bottom panel of Fig. 1, {and taking into account the errors.}

In Fig.~\ref{fig:RCvsModelIC2574} we showed how our model predictions for the case of IC2574 RC. There, the SPARC RC of IC2574 (dots with error-bars) is compared with our simulation's most similar galaxy RC (solid line). The plot also displays the contribution to the RC coming from stars, as the dotted green line, from the gas disk, the dashed red line, and from the total baryonic mass with the blue line. It shows a good agreement between our calculated RC and the SPARC's. {This is also the case for other galaxy properties, such as the baryonic mass profile (star, and gas), and half-mass radius, almost equal to 5kpc, in very good agreement with the observations (check http://astroweb.cwru.edu/SPARC/).}

The case of UCG05750 was not plotted since the RC fit using {the same correction gives as good results as for IC2574.}

\section{Conclusions}\label{sec:Conclusions}

In this paper, we studied the problem of the diversity of RCs shapes in dwarf galaxies \citep{Oman2015}. 
To this aim, we simulated 100 galaxies with similar characteristics to a subsample of the SPARC compilation, and compared the distributions of galaxies of that subsample in the $V_{\rm 2 kpc}$-$V_{\rm Rlast }$ plane with the 
simulated galaxies'. The distributions scatter, and trend show good agreement between the SPARC compilation and our galaxies. However, two outliers are present.
Determining the circular velocity in the galactic plane, instead of from the spherical symmetrical evaluation, we showed that the two galaxies
are no longer outliers. We also showed how the prediction of our model is in agreement with the observations for one of the outliers, namely IC2574. 
~\\

\section*{Acknowledgments}

\acknowledgments{ADP was supported by the Chinese Academy of Sciences and by the President's International Fellowship Initiative, grant no. 2017 VMA0044. MLeD acknowledges the financial support by Lanzhou University starting fund.}
~\\


\externalbibliography{yes}
\bibliography{old_MasterBib1}









\end{document}